\documentclass{aa}
\usepackage[dvips]{graphicx}
\begin{document}

%
   \title{Volume Filling Factors of the ISM Phases in Star Forming Galaxies}
\subtitle{I. The R\^ole of the Disk-Halo Interaction}

   \author{Miguel A. de Avillez\inst{1}
          \and
          Dieter Breitschwerdt\inst{2,3}
          }
   \offprints{M. A. de Avillez}

   \institute{Department of Mathematics, University of \'Evora,
              R. Rom\~ao Ramalho 59, 7000 \'Evora, Portugal \\
              email: mavillez@galaxy.lca.uevora.pt
     \and
        Max-Planck-Institut f\"ur Extraterrestrische Physik,
        Giessenbachstra{\ss}e, Postfach 1312, D-85741 Garching, Germany\\
        email: breitsch@mpe.mpg.de
   \and Present address: Institut f\"ur Astronomie, Universit\"at Wien, 
        T\"urkenschanzstr. 17, A-1180 Wien, Austria\\
        email: breitschwerdt@astro.univie.ac.at
            }

   \date{Received mm dd, year; accepted mm dd, year}

   \titlerunning{Volume Filling Factors of the ISM Phases}
   
   \abstract{The r\^ole of matter circulation between the disk and
     halo in establishing the volume filling factors of the different
     ISM phases in the Galactic disk (${\mathbf |z|\leq 250}$ pc) is
     investigated, using a modified version of the three-dimensional
     supernova-driven ISM model of Avillez (2000).  We carried out
     adaptive mesh refinement simulations of the ISM with five
     supernova rates (in units of the Galactic value),
     $\sigma/\sigma_{Gal}=1$, 2, 4, 8 and 16 (corresponding to
     starburst conditions) using three finer level resolutions of 2.5,
     1.25 and 0.625 pc, allowing us to understand how resolution would
     affect the volumes of gas phases in pressure equilibrium. We find
     that the volume filling factors of the different ISM phases
     depend sensitively on the existence of a duty cycle between the
     disk and halo acting as a pressure release valve for the hot ($T>
     10^{5.5}$ K) phase in the disk. The mean occupation fraction of
     the hot phase varies from about 17\% for the Galactic SN rate to
     $\sim 28$\%, for $\sigma/\sigma_{Gal}=4$, and to $44\%$ for
     $\sigma/\sigma_{Gal}=16$. The amount of cold gas (defined as the
     gas with $T<10^{3}$ K) picked up in the simulations varies from a
     value of 19\% for $\sigma/\sigma_{Gal}=1$ to $\sim 5$\% for
     $\sigma/\sigma_{Gal}=4$ and $\leq 1$\% for higher SN rates.
     Background heating prevents the cold gas from immediate collapse
     and thus ensures the stability of the cold gas phase. The warm
     phase has volume filling factors varying between $0.25$ and
     $0.37$ for the three lowest SN rates used in the simulations.
     Overall the filling factor of the hot gas does not increase much
     as we move towards higher SN rates, following a power law of
     $\langle f_{\mbox{v, hot}}\rangle \propto
     \left(\sigma/\sigma_{Gal}\right)^{0.363}$. Such a modest
     dependence on the SN rate is a consequence of the evacuation of
     the hot phase into the halo through the duty cycle. This leads to
     volume filling factors of the hot phase considerably smaller than
     those predicted in the three-phase model of McKee \& Ostriker
     (1977) even in the absence of magnetic fields.
     \keywords{hydrodynamics -- galaxies: ISM -- galaxies: kinematics
       and dynamics -- Galaxy: disk -- Galaxy: evolution -- ISM:
       bubbles -- ISM: general -- ISM: kinematics and dynamics -- ISM:
       structure}}
\maketitle

\section{Introduction}
In their seminal paper of a three-phase model regulated by supernova
explosions in an inhomogeneous medium, McKee \& Ostriker (1977)
predicted a volume filling factor of the hot intercloud medium (HIM)
of $f_{v, hot} = 0.7 - 0.8$.  However, observations point to a
value of $\sim 0.5$ (e.g., Dettmar 1992) or even lower when external
galaxies are taken into account (e.g.,\ Brinks \& Bajaja 1986).  A way
out has been suggested by Norman \& Ikeuchi (1989) by the so-called
chimney model, in which hot gas can escape into the halo through
tunnels excavated by clustered supernova (SN) explosions. Indeed X-ray
observations of several nearby edge-on galaxies have revealed
extended, galaxy-sized halos (e.g., Wang et al. 2001). The transport
of gas into the halo is, however, still controversial, and arguments
that superbubble (SB) break-out may be inhibited by a large-scale disk
parallel magnetic field have been put forward by several authors (e.g.,
Mineshige et al. 1993).

It is true that magnetic tension forces can in principle considerably
decelerate an upward flow, so that it will eventually stall before
break-out. Tomisaka (1998) has performed 3D MHD simulations of
expanding superbubbles including radiative cooling.  He finds that
bubble confinement only occurs when the energy injection rate is below
a critical value of $\sim 10^{37} \, {\rm erg} \, {\rm s}^{-1}$ (see
also MacLow \& McCray 1988) and/or the scale height of the field is
infinite. Coupling the field distribution to the vertical gas density
distribution, as seems more realistic for frozen-in field lines, e.g.
$B \propto \rho^{1/2}$, break-out will occur even if the midplane
field strength is as high as $5 \mu {\rm G}$.

Attempts to determine the occupation fraction of the different
phases, and in particular of the hot gas, by means of modelling
the effects of SNe and SBs in the ISM, have been carried out by
several authors (e.g.,\ Ferri\`ere 1995, 1998; Korpi et al. 1999).
However, these models do not include the circulation of gas
between the disk and the \emph{full} halo, thus being unable to
resolve the high-$z$ region; neither do they take into account the
mixing between the different phases. Therefore, an estimate of the
volume filling factors may be misleading.

Large scale simulations of the ISM in 2D (Rosen \& Bregman 1995)
and in 3D (Avillez 2000; Avillez \& Berry 2001) show the
importance of the disk-halo interaction in the dynamics and
evolution of the ISM in disk galaxies, which is intimately related
to the vertical structure of the thick gas disk and to the rate of
occurrence of supernovae per unit area in the Galactic disk. A
major consequence of this interaction is the establishment of a
duty cycle between the disk and halo by hot gas breaking through
the thick gas disk, either in a violent way through chimneys or in
a secular fashion through the buoyant rising of hot gas (Avillez
2000; Avillez \& Mac Low 2001), into the upper parts of the thick
gas disk (the disk-halo interface). From here it escapes into the
halo setting up a Galactic fountain (some of this gas may escape
from the Galaxy as a wind, cf.~Breitschwerdt et al. 1993) whose
major fraction returns to the disk as cold gas (after condensing
into clouds), interacting with the thin gas disk.

In this paper we investigate how the volume filling factors of the
different ISM phases in the Galactic disk (which we consider hereafter
to be the region with $-250 \leq z\leq 250$ pc) evolve with the local
Galactic SN rate (varying from the present value to a factor of 16 times 
higher) over a sufficiently long time scale in order to set
up a dynamical equilibrium condition.  We show that matter circulation
\emph{including the Galactic halo} is an important ingredient in
explaining the volume filling factors of the stable phases in the ISM,
in particular a much lower value for the hot intercloud medium (HIM),
even without magnetic fields, is obtained than advocated by McKee \&
Ostriker (1977). The present study makes use of a modified version of
the supernova-driven ISM model of Avillez (2000) and carries out
simulations on grids of kpc-scale size with a spatial resolution as
fine as 1.25 pc and 0.625 pc, respectively, in order to study the ISM
in the disk and halo of star forming galaxies within a wide range of
different supernova rates. The simulation time scales are of the order
of 400 Myr so that the disk-halo-disk cycle can be fully established,
thus allowing the system to completely lose its memory of the initial
conditions.

In \S2 we review the three-dimensional SN-driven ISM model used in the
present work and the numerical schemes applied to it, as well as the
set-up of the current study. In \S3, we describe the global evolution
of the ISM as seen in these runs. Section 4 discusses the results and
their implications.  Finally \S5 presents a summary of the main
results and final remarks.

\section{SN-Driven ISM modelling}
\label{snmod}

\subsection{Numerics}

In the current work we report on large-scale simulations of the
ISM including the disk and Galactic halo using a modified version
of the SN driven ISM model of Avillez (2000) coupled to a
three-dimensional HD code using adaptive mesh refinement (AMR) in
a block-based structure, that relies on virtual topologies of CPUs
created through Message Passing Interface (MPI) calls. This
topology is associated to the computational domain that is divided
into blocks, each having N=n$_{x}\times$n$_{y}\times$n$_{z}$
cells. When a refinement is required, a block is split into eight
new blocks (children) of N cells. This process is repeated until
the finest level of resolution is reached. All the information
relative to the tree structure is preserved in the virtual
topology, being only necessary to query the different CPUs to
learn their location in this topology, and therefore, the location
of their neighbors, children and parents. Each grid block is
refined periodically in regions where steep pressure gradients
appear. The local increase of the number of cells corresponds to
an increase in linear resolution by a factor of two. The adaptive
mesh refinement scheme is based on Berger \& Colella (1989), but
the grid generation procedure follows that described in Bell et
al.\ (1994). The gas dynamics part of the code uses the
piecewise-parabolic method of Colella \& Woodward (1984), a
third-order scheme based on a Godunov method implemented in a
dimensionally-split manner (Strang 1968) that relies on solutions
of the Riemann problem in each zone rather than on artificial
viscosity to follow shocks.

\subsection{Model}
\subsubsection{Setup of basic processes}

The present model introduces substantial improvements to that of
Avillez (2000), namely: (i) inclusion of background heating due to
starlight varying with $z$ (Wolfire et al. 1995) and kept constant in
the directions parallel to the Galactic plane ($z=0$) where it is
chosen to initially balance radiative cooling at 9000 K. With the
inclusion of background heating the gas at $T<10^{4}$ K becomes
thermally bistable; (ii) use of a tabulated cooling function taken
from Dalgarno \& McCray (1972) with an ionization fraction of 0.1 at
temperatures below $10^4$~K and a temperature cutoff at 10~K; (iii)
inclusion of SNe type Ia with a scale height of 325 pc (Freeman 1987);
(iv) OB stars form in regions with a density and temperature
thresholds of 10 cm$^{-3}$ and 100 K, respectively, and their number,
masses and main sequence life times $\tau_{MS}$ are determined from
the initial mass function. Roughly sixty percent of the OB stars
explode within the cluster, while the remaining (composed of the
lowest mass stars whose location is determined kinematically by
attributing to each star a random direction and a velocity) explode in
the field with a scale height of 90 pc. The associations form in a
layer with a scale height of 46 pc. The time interval between the
explosions of all OB stars is determined by their $\tau_{MS}$ for the
given SN rate.

As in the Avillez (2000), the present model includes the fixed
vertical gravitational field provided by the stars in the disk and the
interstellar gas is initially setup with a density stratification
distribution that includes the cold, cool, warm, ionized and hot gases
as described in Ferri\`ere (1998).  This ignores, by the way,
curvature terms in the perpendicular $r-z$-plane, that may become
relevant at some distance from the plane and induce radial motions. On
the other hand, this effect will be small as long as the flow remains
supersonic, since the gas has virtually no time to establish lateral
pressure equilibrium.  The model allows for the motion of the
associations and includes magnetic fields and cosmic rays. The latter
two are the subject of forthcoming papers.

\subsubsection{Further processes}

Owing to the observational complexity of the ISM and the nonlinearity
of the dominant processes occurring there, we have set up a model with
the most important ingredients, neglecting other processes that may be
relevant too, like differential rotation in the case of extended
horizontal gas flows, self-gravity in case of Jeans unstable clouds,
the gravitational field of a dark matter halo component, thermal
conduction, and, most importantly, magnetic fields.  The latter is the
subject of a forthcoming paper (Avillez \& Breitschwerdt 2004).
Self-gravity is neglected, since with the cooling function we are
using, we are not able to treat the formation of molecular clouds
(also dust cooling is not included). In this respect our calculations
are incomplete. However, it is known that the molecular phase
decouples due to gravitational instability, and hence is not even in
rough pressure equilibrium with the other ISM phases.

The reason for attacking the problem in this way, in contrast to other
authors who try to implement as many processes as possible with
computing power being the major limitation, is the following.  We
believe that in order to gain insight into the nonlinear physics of a
complex system like the ISM, driven by temporally and spatially
variable energy input, we should first try to understand the interplay
and effects of the most dominant processes.  Therefore we do not
strive for completeness here, but for a better understanding of the
ISM gas dynamics. As a brief comparison with the observations shows,
and, as will be shown in more detail in the second paper, the
quantitative values that we obtain for the volume filling factors of
the ISM gas are already in remarkably good agreement with the data.
Below we discuss in some detail the relevance of processes that are
known to be important in the ISM, but have been neglected in the
present simulations.

\begin{itemize}

\item  Thermal conduction: The importance of
  thermal conduction in the ISM has been the subject of debate for at
  least 25 years. There is some agreement that Spitzer conductivity
  does certainly overestimate the heat flux by a large amount, since
  the inclusion of an even weak magnetic field reduces the mean free
  path of the electrons to a gyroradius, which is orders of magnitude
  below any astrophysical length scale. Saturated conduction (e.g.\
  Cowie \& McKee 1977) is an improvement, when steep temperature
  gradients occur. In this case the heat flux will be independent of
  the temperature gradient. But the flux would be reduced by
  $\cos\theta$, where $\theta$ is the angle between the local magnetic
  field direction and the temperature gradient. It has been argued
  recently in the context of cooling flows in clusters of galaxies
  (e.g.\ Narayan \& Medvedev 2001), that turbulence may boost the
  conduction rate and thus overcome the suppression by a magnetic
  field. However, as has been pointed out by Malyshkin \& Kulsrud
  (2001), there are two effects that inhibit heat flow even in the
  case of heat conduction in stochastic magnetic fields: (1) the
  electrons have to spiral along the lines of force, which are highly
  tangled, and therefore have to drift a long way, encountering on
  average smaller temperature gradients, (2) the electrons may become
  temporarily trapped on their journey by magnetic mirrors, until
  released by decreasing their pitch angle sufficiently by collisions.
  
  We have not included heat conduction in the present paper, since as
  a second order process it should be slower than hydrodynamic mixing.
  As is shown in this paper, supernova induced turbulence is quite
  strong in the ISM, and therfore we think that turbulent diffusion,
  induced by turbulent motions, will be the more efficient mixing
  process.
  
\item Coriolis forces: The present set-up ignores Galactic rotation,
  which can give rise to radial motions. A measure of the importance
  of Coriolis forces with respect to inertial forces is the Rossby
  number $\epsilon_R = v/(\Omega_0 L) \ll 1$, where $L \sim 1 $ kpc is
  the maximum size of our grid. As far as the hot medium is concerned
  the rms velocity component is of the order of 100 km/s and hence
  $\epsilon_R \sim 4$, for $\Omega_0 = 8.4 \times 10^{-16} \, {\rm
    s}^{-1}$.  Moreover, the hot gas breaking out of the disk
  following the density gradient has a dominant $z$-component, for
  which the Coriolis term vanishes.
  
\item Dark matter halo: Up to a distance of $\sim 10$ kpc the amount
  of dark matter pulling the gas down is small compared to the disk
  potential. To see this, one can look at the gravitational potential,
  for which in case of the Milky Way one can use for the bulge and
  disk component the mass model of Miyamoto \& Nagai (1975) plus a
  spherical dark matter halo component as suggested by Innanen (1973).
  Comparing the gravitational acceleration terms near the solar
  circle, $g_{\rm disk}$ and $g_{\rm halo}$ of the bulge plus disk and
  the dark matter halo, respectively, for a fluid element at vertical
  distances $z=5$ and $z=10$ kpc, one obtains [$g_{\rm disk}(z=5) =
  1.19 \times 10^{-8}$, $g_{\rm halo}(z=5) = 1.23 \times 10^{-9}$] and
  [$g_{\rm disk}(z=10) = 8.72 \times 10^{-9}$, $g_{\rm halo}(z=10) =
  1.96 \times 10^{-9}$], in CGS units respectively.  Only at $z
  \approx 35$ kpc do the two terms become comparable, with the halo
  component dominating at larger heights. Thus neglecting the dark
  matter component in the simulations, where $z \leq 10$ kpc, is fully
  justified for spiral galaxies.

\end{itemize} 

\begin{figure*}[thbp]
  \centering
\vspace*{1.2in}
Left panel: mavillez$\_$fig1a, Middle panel: mavillez$\_$fig1b, Right panel: mavillez$\_$fig1c
\vspace*{1.2in}

\vspace*{1.2in}
Left panel: mavillez$\_$fig1d, Middle panel: mavillez$\_$fig1e, Right panel: mavillez$\_$fig1f
\vspace*{1.2in}

\vspace*{1.2in}
Left panel: mavillez$\_$fig1g, Middle panel: mavillez$\_$fig1h, Right panel: mavillez$\_$fig1i
\vspace*{1.2in}

\caption{Two dimensional cuts, through the 3D data cube, showing the $\rho/\rho_{0}$ (first column), $P/k$ (middle column) and $T$ (right column) distribution in the Galactic plane for $\sigma/\sigma_{Gal}=1$ (first row), 2 (second row) and 4 (bottom row).
\label{1.25pc-GP}}
\end{figure*}

\subsection{Simulations}
The simulations use a computational domain that contains a section of
the Galaxy with an area of 1 kpc$^{2}$ and a vertical extension from
-10 to 10 kpc. The innermost edge lies 8.5 kpc from the Galactic
centre, coinciding with the solar circle. The grid resolution is 10 pc
except in the layer between -500 and 500 pc, where two, three and four
levels of AMR are used, yielding a finest level resolution of 2.5,
1.25 and 0.625 pc, respectively. For the 2.5 and 1.25 pc resolutions
five runs using $\sigma/\sigma_{Gal}=1$, 2, 4, 8 and 16 were carried
out, while the 0.625 pc runs used $\sigma/\sigma_{Gal}=1$, 2 and 4. At
the cap surfaces $z=-10$ kpc and $z=10$ kpc outflow boundary
conditions were imposed and periodic ones along the four vertical
boundary surfaces. The simulations were evolved for 400 Myr.  Such a
large grid size and long evolution time scale is needed to follow the
time-dependent evolution of the ISM allowing several generations of
young massive stars to occur within the computational domain. The
highest resolution simulations are used to answer the crucial
question: does each refinement of the grid lead to a qualitatively and
quantitatively different picture of the ISM or does the distribution
of gas over the various phases converge to values depending mainly on
the SN rate and \emph{not} on grid \emph{resolution}?

\section{Results}
\subsection{Global evolution of the ISM}

The simulations start with a set of hydrodynamic and thermodynamic
variables $\rho$, $P_{gas}$ and $T$ taken from observations.  However,
the initially stratified distribution does not hold for long as a
result of the lack of equilibrium between gravity and (thermal,
kinetic and turbulent) pressure during the ''switch-on'' phase of SN
activity. As a consequence the gas distributions in the upper ($z=10$
kpc) and bottom ($z=-10$ kpc) parts of the grid collapse into the
midplane, leaving low density material behind.  When enough supernovae
have gone off in the disk building up the required pressure support
most of the collapsed gas expands supersonically through the grid
filling all the computational domain. A continuous flow between the
disk and halo is then set up, where the upward and downward flowing
gas come into some sort of dynamical equilibrium.

The gas in the disk also reaches equilibrium, although on a different
time scale, mainly determined by the input of energy into the ISM by
SNe, diffuse heating and the energy lost by adiabatic (due to
expansion of SNRs, SBs and escaping disk gas) and radiative cooling.
The disk equilibrium is related to the time scale that pressure needs
to build up there as a result of SN explosions, while on the global
halo scale such an equilibrium is only possible after the full
establishment of the duty cycle of the warm and hot gas and its
circulation between the disk and the halo, which takes several hundred
Myr. An upper limit for this can be estimated by calculating the flow
time, $\tau_f$, that the gas needs to travel to the critical point of
the flow in a steady-state (see Kahn 1981). This is the characteristic
distance from which information in a thermally driven flow can be
communicated back to the sources.  Then $\tau_f \sim r_c/c_s$, where
$r_c$ and $c_s$ are the location of the critical point and the speed
of sound, respectively. For spherical geometry, the critical point can
be simply obtained from the steady state fluid equations, $r_c \sim G
M_{gal}/(2 c_s^2)$, which yields with a Milky Way mass of $M_{gal}
\approx 4 \times 10^{11} \, {\rm M_{\odot}}$ a distance $r_c \approx
31.3$ kpc for an isothermal gas at $T= 2 \times 10^6$ K (corresponding
to a sound speed of $1.67 \times 10^7 {\rm cm} \, {\rm s}^{-1}$), and
thus $\tau_f \sim 180$ Myr as an upper limit for the flow time. For
comparison, the radiative cooling time of the gas at a typical density
of $n=2 \times 10^{-3} \, {\rm cm}^{-3}$ is roughly $\tau_c \sim 3 k_B
T/(n \Lambda) \approx 155$ Myr for a standard collisional ionization
equilibrium cooling function $\Lambda = 8.5 \times 10^{-23} \, {\rm
  erg}\, {\rm cm}^3 \, {\rm s}^{-1}$ of gas with cosmic abundances.
This value is apparently of the same order as the flow time, ensuring
that the flow will not only cool by adiabatic expansion, but also
radiatively, thus giving rise to the fountain return flow, which is
the part of the outflow that loses pressure support from below and
therefore cannot escape. Note that $r_c$ is the maximum
extension of the fountain, more than an order of magnitude larger,
than in simulations with grids with a vertical extension of 1 kpc
above and below the midplane. Those calculations definitely miss an
important component of the galactic gas dynamics.

In the disk, on the other hand, the cooling time scale is much shorter
than in the halo, because of the higher ISM densities there.  We thus
find from our simulations that typical time scales of pressure
fluctuations about a mean value of $P/k_B \sim 2600 \, {\rm cm}^{-3}
\, {\rm K}$ are of the order of $30$ Myr, which is by the way also a
typical time scale for superbubble evolution.

However, it should be emphasized, that since disk and halo are coupled
dynamically  not only by the escape of hot gas, but also by the
fountain return flow striking the disk, the disk equilibrium
will suffer secular variations on time scales of the
order of $100 - 200$ Myr.

Cuts through the 3D data cube at $z=0$ pc and taken at 400 Myr
presented in Figure~\ref{1.25pc-GP} show the density (left
column), ${\rm P/k}$ (middle column) and temperature (right column)
distributions in the Galactic plane at 400 Myr of evolution, well
after the system has reached dynamical equilibrium. The supernova
rates in units of the Galactic value are: 1, 2, and 4, and the
resolution of the finest AMR level is 1.25 pc. These images show
the structure and morphology of the different ISM phases defined
here as: cold ($T<10^{3}$~K), cool ($10^{3}<T<10^{4}$~K), warm
($10^{4}<T<10^{5.5}$~K) and hot ($T>10^{5.5}$~K). The morphology
of the ISM and the volume filling factors of the various phases
obviously vary with SN rate. Not surprisingly, the amount of cold
gas reduces with increasing SN rate, while the amount of the warm
and hot gas increases. The images also show the expansion of SNRs
and the interaction of shock waves, which can be best observed in
the pressure maps where the shocks are represented by high
pressure and thin structures in red. This is not seen in the
density or temperature maps. A comparison between these maps shows
that most of these shocks are located at the interfaces between
hot and intermediate temperature gas, i.e., $10^{4}<T<10^{5.5}$ K.

Inspection of the pressure distribution shows that the most abundant
colour is yellow-green, implying an average ${\rm P}/{\rm k}
\sim 2600 \, {\rm K} \, {\rm cm}^{-3}$ (see also
Fig.~\ref{temppdfs-1.25pc}, upper right panel) for $\sigma =
\sigma_{Gal}$, and about $10,000$ for $\sigma = 4 \sigma_{Gal}$. This
is in agreement with the observational fact that the cool, warm and
hot phase are in rough pressure equilibrium. This should however not
be confused with a $local$ equilibrium. On the contrary, severe
deviations from the equilibrium values can occur locally due to SN
activity and thermal instabilities (see the red an blue regions in
Fig.~\ref{1.25pc-GP}, middle column).

For the Galactic SN rate, once the system reaches its equilibrium
configuration, a thin disk of cold gas forms in the Galactic
plane, and, above and below, a thick inhomogeneous gas disk forms
(this vertical structure of the disk is similar to that found in
Avillez 2000). The code does not explicitly follow ionization
states, but we can trace cool gas with a temperature $T\leq
10^{4}$ K and a scale height of 180~pc, and warm gas with
$10^{4}\le T\le 10^{5}$ K and a scale height of 1~kpc. These
distributions reproduce those described in Dickey \& Lockman
(1990) and Reynolds (1987), respectively. The thick gas disk is
punctured by chimneys that result from superbubbles occurring on
either side of the midplane at a height greater than $100$ pc. As
they grow, they elongate along the $z$-direction, owing to the
local density stratification of the ISM.  Chimneys in the
simulation typically have widths of approximately 150-200 pc. They
inject high temperature gas directly from the Galactic disk into
the halo, breaking through the cool and warm layers that compose
the thick disk. This hot gas then contributes to the galactic
fountain. Thus, the upper parts of the thick warm disk form the
disk-halo interface, where a large scale fountain is set up by hot
ionized gas escaping in a turbulent convective flow.

For higher SN rates a similar $z$-structure of a thick gas disk is seen,
but with a higher vertical extension. This is mainly due to the rate
of SNe in the field rather than to the clustered SNe, which drive
superbubbles breaking through the thick gas disk and injecting their
matter directly into the halo. The evolution of the vertical structure
of star-forming galaxies with different SN rates will be discussed in
more detail in a forthcoming paper.

\subsection{Probability distribution functions}
\label{pdfs}

\begin{figure}[thbp]
\centering
\vspace*{1.4in}
mavillez$\_$fig2a
\vspace*{1.3in}

\vspace*{1.4in}
mavillez$\_$fig2b 
\vspace*{1.3in}

\vspace*{1.4in}
mavillez$\_$fig2c
\vspace*{1.3in}
\caption{Averaged volume-weighted temperature PDFs over the periods of 0-50 Myr (red) and 350-400 Myr
(black) calculated using 51 snapshots taken at time intervals of 1 Myr. The supernova rates used in
these models are: $\sigma/\sigma_{Gal}=1$ (top panel), 4 (middle panel), and 8 (bottom panel). The
resolution of the finest AMR level is 1.25 pc.
\label{temppdfs-1.25pc}
}
\end{figure}
\begin{figure}[thbp]
\centering
\vspace*{1.4in}
mavillez$\_$fig3a
\vspace*{1.3in}

\vspace*{1.4in}
mavillez$\_$fig3b 
\vspace*{1.3in}

\vspace*{1.4in}
mavillez$\_$fig3c
\vspace*{1.3in}
\caption{Averaged volume-weighted density PDFs over the period
350-400 Myr calculated using 51 snapshots taken at time intervals
of 1 Myr for the SN rates shown in Figure~\ref{temppdfs-1.25pc}.
The finest level AMR resolution is 1.25 pc.
\label{denpdfs1.25pc}
}
\end{figure}

\begin{figure}[thbp]
\centering
\vspace*{1.4in}
mavillez$\_$fig4a
\vspace*{1.3in}

\vspace*{1.4in}
mavillez$\_$fig4b 
\vspace*{1.3in}

\vspace*{1.4in}
mavillez$\_$fig4c
\vspace*{1.3in}
\caption{Averaged volume-weighted pressure PDFs over the period
350-400 Myr calculated using 51 snapshots taken at time intervals
of 1 Myr for the SN rates shown in Figure~\ref{temppdfs-1.25pc}.
The finest level AMR resolution is 1.25 pc. \label{prepdfs1.25pc}
}
\end{figure}
A major consequence of the set-up of the disk-halo cycle by SN activity is
that the system loses any recollection of its initial conditions as
most of the disk gas has already travelled into the halo and came back
to the disk. Thus, the effects of initial conditions are not
present in the temperature probability distribution functions
(PDFs) of the system for different SN rates.

Figure \ref{temppdfs-1.25pc} compares volume weighted histograms of
the temperature over the periods of 0-50 Myr (red) and 350-400 Myr
(black) calculated using 50 snapshots taken at time intervals of 1 Myr
for the runs with $\sigma/\sigma_{Gal}=1$, 4, and 8 and a finest AMR
resolution of 1.25 pc. These PDFs indicate that for low $\sigma$, the
temperature peak is at about 2000 K, making the cold/warm HI gas the
most abundant gas phase, consistent with the density PDFs
(Figure~\ref{denpdfs1.25pc}) and also with observations. This peak is
shifted towards $T \approx 40000$ K, for $\sigma = 16 \,
\sigma_{Gal}$, making the warm phase the most important one.  At the
same time the relative importance of the hot phase increases as well,
and one is moving towards a bimodal distribution. It can be directly
seen from Fig.~\ref{temppdfs-1.25pc} that a simulation time of only 50
Myrs is not sufficient to obtain this result, mainly because the
effect of upwards transport is to establish a duty cycle, which takes
of the order of a few hundred Myrs. In any of the cases shown in the
figures the averaged PDFs for the initial 50 Myr have two pronounced
peaks, one around 8000 K and the other around $5\times 10^{6}$ K. Note
that with the increase of SN rate the PDFs of the first 50 Myrs suffer
large variations, indicating that the loss of recollection of the
initial conditions will occur earlier for the highest SN rates,
suggesting that in a time-asymptotic sense the SN rate is the
controlling parameter.
The corresponding averaged volume-weighted PDFs of the density and
pressure distribution in the Galactic disk over the period 350-400 Myr
are shown in Figures~\ref{denpdfs1.25pc} and \ref{prepdfs1.25pc}. It
can be seen that there are \emph{three} distinct peaks in the density
PDFs corresponding to the three most abundant regimes (cool, warm and
hot). The cold regime has a peak that decreases steeply with increase
of SN rate. These peaks have similar pressure ranges. The pressure
PDFs show that the range of the total pressure (dashed lines)
decreases slightly with increase of the SN rate. For $\sigma =
\sigma_{Gal}$ and $<$dN/N$>=10^{-2}$ , total pressure spans three
orders of magnitude from $10^{-14}$ to $10^{-11}$, while for higher SN
rates of 8 and 16 times the Galactic value the total pressure spans
three and two orders of magnitude, respectively.  This is indicative
of the large variation in the pressure distribution between the
different temperature regimes, and suggests that there are no real
phases, i.e. co-existing thermodynamic regimes with different density
and temperature but in pressure equilibrium.

\subsection{Volume filling factors}
\label{vffs}
At $\sigma = \sigma_{Gal}$, the hot phase has a volume filling factor
$f_{v,hot} \sim 0.17$, comparable to that of the cold gas, and the
warm/cool phase is dominating. Increasing to 4, 8 and 16 times
$\sigma_{Gal}$ pushes the warm gas to take over in volume, while there
is a slight increase of the occupation fraction of the hot gas. It is
interesting to note that the peak of the warm gas still surpasses that
of the hot gas in magnitude even for a factor of 16 (see
Fig.~\ref{VFF-SNrate}).  This must be due to the fact that the
reservoir of cold gas is used up by increased Lyman continuum photon
absorption and shock heating, whereas the hot gas escapes into the
halo.  Between $\sigma = 4$ and $16 \, \sigma_{Gal}$, the cold/cool
phase is reduced substantially and eventually to insignificance. This
must have direct consequences for the formation of molecular clouds
and thus for continuous star formation. It seems plausible that this
leads eventually to self-regulation.

Figure~\ref{vff} shows the history of of the volume filling factors of
the different phases in the simulated disk for
$\sigma/\sigma_{Gal}=1$, 4, and 8, while Table~\ref{Avervff} and
Figure~\ref{VFF-SNrate} show the variation with supernova rate of the
time averaged volume filling factors of the different phases, over the
period 300-400 Myr calculated using 101 snapshots with a time interval
of 1 Myr.

The distribution of the occupation fraction of the different phases
comes to an equilibrium at about 100 Myr for $\sigma/\sigma_{Gal}>1$,
while for the Galactic SN rate it occurs somewhat later, i.e.\ at
about 200 Myr. This is a result of the set-up of the disk-halo cycle,
which for the Galactic SN rate takes about 200 Myr to establish, while
for higher SN rates it takes less time in evacuating the hot gas from
the disk. This is presumably due to an over-pressure with respect to
the ambient medium and a higher sound speed of the injected hot gas.
But in any case, the timescale for the cycle to be completed, even
with higher SN rates, is always larger than $100$ Myr, thus ruling out
simulations with lower evolution times.

\begin{figure}[thbp]
\centering
\vspace*{1.4in}
mavillez$\_$fig5a
\vspace*{1.3in}

\vspace*{1.4in}
mavillez$\_$fig5b 
\vspace*{1.3in}

\vspace*{1.4in}
mavillez$\_$fig5c
\vspace*{1.3in}
\caption{Time evolution of the volume filling factors of the cold
 ($T\leq10^{3}$ K), cool ($10^{3}<T\leq10^{4}$ K), warm
($10^{4}<T\leq10^{5.5}$ K), and hot ($T>10^{5.5}$ K) phases for
$\sigma/\sigma_{Gal}=1$ (top panel), 4 (middle panel), and 8 (bottom panel).
The finest AMR level resolution is 1.25 pc. Note that the line for the cold gas is slightly over $f_{v}=0$, for $\sigma/\sigma_{Gal}=8$.
\label{vff} }
\end{figure}

Most remarkably, for $\sigma/\sigma_{Gal}=1$, the hot ($T>10^{5.5}$ K)
gas has a moderately low volume filling factor (it fluctuates around 0.17)
in agreement with observations ($\sim 20$\%) even in the absence of
magnetic fields, and is mainly distributed in an interconnected tunnel
network, and in some cases it is even confined to isolated bubbles as
seen in Figure~\ref{1.25pc-GP}. With the increase of the supernova
rate to four and eight times the galactic rate the occupation volume
of the hot gas increases to about $28$\% and $35$\%, respectively,
after 400 Myr of disk evolution, which is still below the predictions
of McKee \& Ostriker (1977). Even for $\sigma/\sigma_{Gal}=16$,
corresponding already to starburst conditions, the volume fraction of
the hot gas increases only to $44$\% after 400 Myr of evolution. Such
a behaviour can be approximated analytically by a power law fit
(Figure~\ref{VFF-SNrate}
\begin{equation}
\langle f_{v,
hot}\rangle=0.16\left(\frac{\sigma}{\sigma_{Gal}}\right)^{0.363} \,,
\end{equation}
with a a RMS percent error of 0.01.

\begin{table}[thbp]
\centering \caption{Average volume filling factors of the
different ISM phases for variable SN rate. The average was calculated
using 101 snapshots (of the 1.25 resolution runs) between 300 and 400
Myr of system evolution with a time interval of 1 Myr. \label{Avervff}
}

\begin{tabular}{ccccc}
\hline
\hline
$\sigma$$^a$ & $\langle f_{v, cold}\rangle$$^b$ &
$\langle f_{v, cool}\rangle$$^c$ &  $\langle f_{v, warm}\rangle$$^d$ & $\langle f_{v, hot}\rangle$$^e$ \\
\hline
1 & 0.19 & 0.39  & 0.25  & 0.17 \\
2 & 0.16 & 0.34 & 0.31 & 0.19 \\
4 & 0.05 & 0.30 & 0.37 & 0.28 \\
8 & 0.01 & 0.12 & 0.52 & 0.35 \\
16 & 0.0 & 0.02 & 0.54 & 0.44 \\
\hline
\multicolumn{5}{l}{$^a$ SN rate in units of the Galactic SN rate.}\\
\multicolumn{5}{l}{$^b$ $T<10^{3}$ K.}\\
\multicolumn{5}{l}{$^c$ $10^{3} <T\leq 10^{4}$ K.}\\
\multicolumn{5}{l}{$^d$ $10^{4} <T\leq 10^{5.5}$ K.}\\
\multicolumn{5}{l}{$^e$ $T> 10^{5.5}$ K.}\\
\end{tabular}
\end{table}

\begin{figure}[thbp]
\centering
\vspace*{1.4in}
mavillez$\_$fig6
\vspace*{1.4in}
\caption{Variation of the average volume filling factor of $T<10^{3}$
  (circles), $10^{3}< T< 10^{4}$ (squares), $10^{4}< T< 10^{5.5}$
  (diamonds) and $T> 10^{5.5}$ K (stars) phases in the simulated disk
  (for $\left| z\right|<250$ pc). The filling factors were averaged
  using 101 snapshots (of the 1.25 resolution runs) separated by
  1 Myr between 300 and 400 Myr of the disk evolution. Red represents
  a power law fit to $\langle f_{v, hot}\rangle$ with the power of
  0.363.
\label{VFF-SNrate}
}
\end{figure}
\begin{figure}[thbp]
\centering
\vspace*{1.5in}
mavillez$\_$fig7a
\vspace*{1.3in}

\vspace*{1.5in}
mavillez$\_$fig7b 
\vspace*{1.3in}
\caption{Comparison between volume filling factors of cold ($T<10^{3}$ K), cool
($10^{3}<T<10^{4}$ K), warm ($10^{4}<T<10^{5.5}$ K) and hot ($T>10^{5.5}$ K) gas for the finest AMR level grid resolutions of 0.625 pc (black), 1.25 pc (red),
and 2.5 pc (green) for the SN rates $\sigma/\sigma_{Gal}=2$ (top panels),
and 4 (bottom panels).
\label{vff-compare-01}
}
\end{figure}
The warm gas becomes the dominant phase for $\sigma/\sigma_{Gal}>2$.
Between $\sigma/\sigma_{Gal}=8$ and 16 the occupation volume of this
phase has a very small increase from 0.52 to 0.54, as a result of the
conversion of part of the cool gas into this phase. The decrease in
the cool gas is a result of its conversion into the warm and hot
phases. The growth of the hot phase in time is a mild one even with
the large increase in the supernova rate. This is a consequence of the
circulation of matter from the disk into the halo which acts as a
pressure release valve for the disk gas.  As the hot gas is vented
into the halo, there is enough space for the $10^{4}< T< 10^{5.5}$ K
gas to be redistributed in the disk.  Furthermore, as there is
recycling, low temperature gas continues to flow into the disk, thus
contributing to the maintenance of the intermediate phases.

For $\sigma/\sigma_{Gal}=2$ there seems to be a small difference in
the volume occupation of the cool (around 34\%) and warm (around 31\%)
phases, while the occupation fraction of the cold gas differs from
these by at most $15\%$ of the total volume. It is interesting to note
that the volume occupation of the cool gas decreases some $5\%$ with
the doubling of the SN rate for $\sigma/\sigma_{Gal}\leq 4$. On the
other hand, there is a reduction of almost $70\%$ of the filling
factor of the cold gas when the supernova rate is increased to
$\sigma/\sigma_{Gal}=4$. This fraction reduces to less than $1\%$ at
$\sigma/\sigma_{Gal}=8$, disappearing completely for higher SN rates.

\subsection{Resolution effects}

Our simulations show how crucial spatial resolution is in order to
capture small scale structures and promote the mixing of different
fluid elements. For instance the amount of gas in the different
temperature regimes, and most importantly in the cold phase, depends
sensitively on spatial grid resolution.  As an example consider the
amount of cold gas, the volume filling factor of which increases for
$\sigma/\sigma_{Gal}=2$ from $\sim 9\%$ in the 2.5 pc resolution
calculations to $\sim 17\%$ in the 1.25 pc and 0.625 pc resolution
cases (s.~Figure~\ref{vff-compare-01}); and the discrepancy for the
cold remains still large, when one increases the supernova rate to
$\sigma/\sigma_{Gal}=4$. For the hot phase, on the other hand, there
is not much difference in $f_{v,hot}$ with increasing resolution. The
largest variation occurs for the smallest SN rates with a reduction of
at most 20\% compared to the values derived from the $\Delta x=2.5$ pc
resolution simulations. The resolution increase also affects the cool
and warm phases in a similar way with an increase in $f_{v,cool}$ and
a decrease in $f_{v,warm}$ for $\sigma/\sigma_{Gal}>2$.

The variations in the volume filling factors are easily understood if
one takes into account that with an increase of resolution it becomes
possible to resolve the smallest scale structures. Thus instead of
averaging out the gas density over larger cells, and thereby wiping
out density peaks, radiative cooling as a nonlinear process can become
more efficient, since high density regions contribute more to the
energy loss rate than low density regions can compensate by an
accordingly lower rate. Since cooling is most efficient for dense gas,
the cool phase is affected most. In addition, the spatial resolution
of shear layers and contact surfaces, gives rise to an increased level
of turbulence and a larger number of mixing layers.  The latter is
most important, because it allows for a faster mixing between parcels
of gas with different temperatures (conduction or diffusion processes
being of second order and hence inherently slow in nature). The small
scale mixing in these simulations is promoted by numerical rather than
molecular diffusion, and therefore, the time scales for mixing in the
different phases to occur is somewhat smaller (because it happens on
larger scales) than those predicted by molecular diffusion theory
(e.g., Avillez \& Mac Low 2002). However, as already mentioned,
turbulent diffusion, as a consequence of the onset of turbulence due
to shear flows, will be most efficient.
\begin{figure*}[thbp]
\centering
\vspace*{2in}
Left panel: mavillez$\_$fig8a, Middle panel: mavillez$\_$fig8b, Right panel: mavillez$\_$fig8c
\vspace*{2in}
\caption{Comparison between grid resolutions of 0.625 pc (black),
1.25 pc (red) and 2.5 pc (green) of maximum density (top panel)
and minimum temperature (bottom panel) for $\sigma/\sigma_{Gal}=1$
(left column), 2 (middle column), and 4 (right column). Note that
in any of the panels there is  large difference between the 2.5
and 1.25 pc resolutions, while the differences between 0.625 pc
and 1.25 pc resolutions are much smaller. This is an indication of
the convergence of the simulations in reproducing the physical
processes involved in the dynamics and evolution of the ISM, thus
representing a fairly realistic distribution of gas in
density-temperature space. \label{dmax01-02} }
\end{figure*}

A comparison between the maximum density, n$_{max}$, and minimum
temperature, T$_{min}$, measured at the different finer level
resolutions reveals that an increase in resolution from $\Delta x=2.5$
to 1.25 pc implies an average increase in n$_{max}$ and a decrease in
T$_{min}$ by factors greater than 5 at any SN rate (cf.\ 
Figure~\ref{dmax01-02}). When resolution is increased from 1.25 to
0.625 pc the differences between n$_{max}$ and T$_{min}$ for all the
SN rates are small and diminish with the increase of SN rate. The high
T$_{min}$ for $\sigma/\sigma_{Gal}=4$ at any resolution is due to the
higher cooling time of the gas supply with on average higher
temperature. Figure~\ref{resolution} compares the average values of
T$_{min}$ and n$_{max}$ calculated between 200 and 400 Myr of
evolution for the three resolutions (for details, see figure caption).

\begin{figure*}[thbp]
\centering
\vspace*{2in}
Left panel: mavillez$\_$fig9a, Right panel: mavillez$\_$fig9b
\vspace*{2in}
\caption{Comparison between the average minimum temperature (left panel) and
  maximum density (right panel) as function of resolution for the
  three SN rates: $\sigma/\sigma_{Gal}=1$ (open circles),
  $\sigma/\sigma_{Gal}=2$ (stars), and $\sigma/\sigma_{Gal}=4$
  (triangles). The plots also show exponential fits to the data
  points. These average values were calculated during the last 200 Myr
  of evolution, i.e. after establishing dynamical equilibrium, so that
  their history does not reflect any memory of the initial conditions.
\label{resolution}
}
\end{figure*}

When a resolution of 0.625 pc is used the average minimum temperature
and maximum density, T$_{min}$ and n$_{max}$, suffer small variations
with respect to those determined at $\Delta x=1.25$ pc simulations for
all the SN rates discussed above. This is a clear indication of the
\textit{convergence} of the simulations in reproducing the physical
processes involved in the dynamics and evolution of the ISM and that
the simulations with $\Delta x\leq 1.25$ pc can represent the real
ISM. From the fit for $\langle$T$_{min} \rangle = 8.35 \exp({\Delta x
  \over 1.106 {\rm pc}})$, and $\langle$n$_{max} \rangle = 442
\exp(-{\Delta x \over 1.46 {\rm pc}})$, respectively, we can see that
there is rapid convergence for $\Delta x\lse 1.1$ pc.

\section{Discussion}
\label{disc} 

The filling factors of some of the phases obtained in this study are
similar to those estimated by Spitzer (1990), Ferri\`ere (1995) and
Avillez (2000), but seem to be in contradiction with predictions of
Ferri\`ere (1998) and are not consistent with Korpi et al. (1999).

The first three authors estimated a volume filling factor of the hot
gas in the Galactic disk of $\sim 0.2$, while for the cold and neutral
phases Ferri\`ere (1995) and Avillez (2000) arrived to similar values.
The largest variation between the volume filling factors in Avillez
(2000) and the present paper is by far in the value of  $f_{v,cold}$, as a
result of the introduction of background heating due to starlight
leading to the formation of thermally stable branches at
$10^{3.9}<$T$<10^{4.2}$ K and T$<100$ K. Most of the cold gas is
located in the unstable branch at temperatures below $10^{3}$ K
(Avillez \& Breitschwerdt 2004).

Based on models which follow the expansion and contraction phases of
SNRs and SBs, Ferri\`ere (1998) estimated the variation of the volume
filling factor of the hot gas with $z$. It is argued that near the
solar circle, the volume occupation of the hot gas increases from
$0.15$ at $z=0$ pc to $0.23$ at $z=200$ pc and decreases gradually for
$z>200$ pc (Figure 12 of Ferri\`ere 1998). The discrepancy between
these results and the calculations reported in this paper can be
understood by the fact that Ferri\`ere's calculations reflect the
distribution of the hot gas \textit{inside} the SNR and SB cavities
whose number decreases with $z$, and therefore the volume filling
factor of hot gas decreases accordingly. This contrasts with the
present paper where in a global 3D simulation the hot gas is not
exclusively connected to individual SNRs or SBs\footnote{In fact we
  see a large number of SNRs and SBs disintegrating on timescales of 1
  to 30 Myrs, respectively, even inside the Galactic disk, where
  density gradients are on average smaller than in the halo.}, but
instead can rise into the halo. As the hot gas breaks out of the {\it
  thick} disk, in which its volume filling factor first decreases, it
eventually starts filling the entire volume of the Galactic halo.
Thus, Ferri\`ere's model does in fact not resolve the vertical gas
transport, and hence misses out on the volume fraction of a
considerable amount of gas, leading effectively to an \textit{increase
} with $z$-height above 1 kpc as described here.

Figure~2 of Korpi et al. (1999) shows the temperature, density and
$P/k$ PDFs for different volumes of the disk: $\left|z\right|<0.25$,
$\left|z\right|<0.5$, and $\left|z\right|<1$ kpc. These are averaged
volume-weighted histograms calculated over a period of 50 Myr using
six snapshots at equal time intervals between 20 and 70 Myr of
evolution. The temperature PDF for the Galactic disk gas has a bimodal
distribution with two strong peaks one at $10^{4}$ K and another at
$\sim 10^{6}$ K. The latter corresponds to a $f_{v,hot}$ varying from
$20-30$\% at $z=0$ pc to $50-60$\% for $z=300$ pc. These values are in
disagreement with those discussed in the present paper for the
Galactic SN rate. The discrepancy is related to the small grid extent
in the $z$-direction (with $-1 \leq z\leq 1$ kpc) used by these
authors, and consequently the increasing escape of material from the
grid with time without any return flow making their model, as the
authors argue themselves, only meaningful for a limited length of
time, which as we discuss below should not exceed 100 Myr.

The need for a duty cycle and the establishment of a global dynamical
equilibrium require the use of an extended grid in the direction
perpendicular to the Galactic plane. The lack of such an extended
$z$-grid inhibits the disk-halo-disk circulation of matter, which
otherwise would return gas to the disk sometime later, with noticeably
increasing effects for the dynamical evolution as time proceeds, which
can be clearly seen in the present simulations.  In grids, which have
a small vertical extension, e.g., only 1 kpc above and below the
midplane\footnote{Note, that the maximum extension of the fountain is
  more than an order of magnitude larger, and therefore such
  restricted calculations definitely miss an important component of
  the galactic gas dynamics.}, compared to the maximum height to which
the hot plasma rises due to the injection of energy and momentum from
the sources, most of the gas escapes from the disk in less than 100
Myr (without ever returning to it). Thus the simulations can only be
followed for a small evolution time, and therefore, the duty cycle
cannot be set up and the system never reaches a dynamical equilibrium
state. Moreover, the volume weighted histograms of the thermodynamic
properties (e.g., density, pressure and temperature) of the disk gas
will retain a memory of the initial evolution. As explained in \S3.1,
after an initial collapse of the matter distribution towards the grid
midplane, pressure is built up and there is a redistribution of matter
on the grid, filling it and allowing a substantial fraction of the gas
to escape through the top and bottom boundaries.  Therefore, averaged
PDFs that including 20 - 50 Myrs snapshots will show the presence of a
large fraction of cold gas from the collapse phase (recall the
averaged 50 Myr PDF shown in the top panel of Figure~2 in \S3.2),
while PDFs constructed at later times will show the dominance of the
hot gas. A combination of these PDFs would give a pronounced bimodal
distribution (even if there is no background heating due to starlight)
corresponding to large volume filling factors for the hot and coolest
ISM phases. In other words such PDFs are a reflection of the
\emph{initial} evolution of the simulations and not of the time when a
dynamical equilibrium is already established.

\section{Summary and final remarks}
\label{conc}

In the present paper we have described a set of 3D hydrodynamical
simulations of the interstellar medium in order to study the
distribution of the ISM phases and how they vary with increasing
star formation rate (i.e.,\ SN rate) in star forming galaxies. The
major goals of this work were (i) to see if the presence of a
disk-halo-disk circulation has a major impact on the volume
filling factors of the hot phase, and (ii) which minimum grid
resolution is needed in order to obtain quantitatively reliable
results that can be compared to observations. The main results of
the present work can be summarized as follows:
\begin{itemize}

\item The occupation fractions of the different ISM phases depend
sensitively on the presence of a duty cycle established between
the disk and halo working as a pressure release valve for the hot
phase.

\item The mean occupation fraction of the hot phase varies from about
  17\% for the Galactic SN rate to 28\% and 44\% for
    $\sigma/\sigma_{Gal}=4$, and 16, respectively. The mean occupation
    fraction follows a power law increase with SN rate, with the power
    law index of $\sim0.363$.

\item The amount of cold gas picked up in the simulations varies
from a value of roughly 19\% (for the Galactic SN rate) to about 5\% for
$\sigma/\sigma_{Gal}=4$ and $\leq 1\%$ for $\sigma/\sigma_{Gal}\geq 8$.

\item The warm phase has occupation fractions varying between $25\%$ and $37\%$ for the three lowest SN rates ($\sigma/\sigma_{Gal}\leq 4$).

\item Background heating is the main reason for the increase in
the amount of cold gas in comparison to that in 3D simulations
without any background heating.

\item A minimum grid resolution of 1.25 pc is needed for quantitatively
reliable results, as the convergence towards the 0.625 pc resolution
simulations shows.

\item A SN rate of $8 - 16$ times the Galactic value already
represents a starburst; there is increasing evidence that most
star forming galaxies have undergone several such phases during
their evolution. In particular for high redshift galaxies,
starbursts seem to have been common. Our simulations show that the
volume filling factor for the hot phase in the disk increases only
moderately from $\langle f_{v, hot} \rangle = 0.35$ and 0.44 for
$\sigma/\sigma_{\rm gal} =8$ and 16, respectively.

\quad This implies that in X-ray observations, the value of $\langle
f_{v, hot}\rangle$ in the \emph{disk} is not a reliable indicator
for starburst. Instead the \emph{size of the halo} in soft X-rays
is strongly correlated with a starburst as can be seen from the
size of the X-ray halos in recent XMM-Newton observations of
\object{NGC 253} (e.g.,\ Pietsch et al. 2001)
and \object{NGC
3079} (Breitschwerdt et al. 2004), and also ROSAT observations of
\object{M 82} (e.g.,\ Bregman et al. 1995) and \object{NGC 253}
(Dahlem et al. 1998, Pietsch et al. 2000).

\item Even for Galactic SN rates the fountain cycle is established and
thus hot gas is present in galactic halos. This explains why also
star forming galaxies with SN rates comparable to the Galaxy exhibit soft
X-ray halos, albeit smaller than in starburst
galaxies, as has been observed for \object{NGC 4631} (Wang et al. 2001), 
which also may have been disturbed by a companion galaxy thereby 
enhancing the star formation rate, and
\object{NGC 891} (Bregman \& Irwin 2002), which is often referred to 
as the twin galaxy to the Milky Way.

\end{itemize}

The calculations presented in this paper do not include the magnetic
field. A parameter study of the effects of the magnetic field and
cosmic rays in the ISM is underway and will be described in Avillez \&
Breitschwerdt (2004). It will deal with the effects of the B-field
dissipation, which is too low for Galactic dynamos (Ferri\`ere 1998),
as well as with the effects of a weak and strong magnetic field. If
the magnetic field is present and is initially mainly oriented
parallel to the disk, transport in the halo may be inhibited, although
not prevented. On larger scales magnetic tension forces become weaker
than on the smallest scales and therefore vertical expansion might
still take place efficiently. Either way the occupation fraction of
the hot gas could be comparable to the values observed in the present
simulations.

\begin{acknowledgements}
  M.A. and D.B are partially supported by the ESO/FCT (Portuguese
  Science foundation) grant PESO/P/PRO/40149/2000. DB acknowledges
  support from the Bundesministerium f\"ur Bildung und Forschung
  (BMBF) by the Deutsches Zentrum f\"ur Luft- und Raumfahrt (DLR)
  under grant 50 OR 0207 and the Max-Planck-Gesellschaft (MPG). The
  0.625 pc simulations discussed in this paper were carried out at
  Compaq Benchmark Center, Toronto, Canada. The authors thank Michael
  Adamson, team leader at Compaq, for all the support and help during
  the time of computing. 
\end{acknowledgements}

\end{document}